# Regge behaviour of structure functions and solution of Dokshitzer-Gribov-Lipatov-Altarelli-Parisi Evolution equations in next-to-leading order at low-x

U. Jamil[1] and J. K. Sarma[2]

Department of Physics, Tezpur University, Napam, Tezpur-784028, Assam, India

**Abstract.** Deuteron and proton structure functions are derived from Dokshitzer-Gribov-Lipatov-Altarelli-Parisi (DGLAP) evolution equations of singlet and non-singlet structure functions in next-to-leading order (NLO) at low-x assuming the Regge behaviour of singlet and non-singlet structure functions at this limit and results are compared with New Muon Collaboration (NMC) and E665 collaboration data sets.

**1 Introduction**

Understanding the low-x behaviour of the structure functions of the nucleon, where x is the Bjorken variable, is interesting both theoretically and phenomenologically. Deep inelastic scattering (DIS) process is one of the most successful experimental methods for the understanding of quark-gluon substructure of hadrons [1-3] from which one gets the measurement of $F_2(x, Q^2)$ (proton, neutron and deuteron) structure functions in the low-x region where $Q^2$ is the four momentum transfer in a DIS process. Structure functions are important inputs in many high energy processes and also important for examination of perturbative quantum chromodynamics (PQCD) [3], the underlying dynamics of quarks and gluons. In PQCD, for high-$Q^2$, the $Q^2$-evolutions of these densities (at fixed-x) are given by the DGLAP evolution equations [4, 5]. The solutions of the DGLAP equations can be calculated either by numerical integration in steps or by taking the moments of the distributions [6]. Among various solutions of this equation, most of the methods are numerical. Mellin moment space [7] with subsequent inversion, Brute force method [8], Laguerre method [9], Matrix method [10] etc. are different methods used to solve DGLAP evolution equations. The shortcomings common to all are the computer time required and decreasing accuracy for $x \to 0$ [10]. More precise approach is the matrix approach to the solution of the DGLAP evolution equations, yet it is also a numerical solution. Thus though numerical solutions are available in the literature, the explorations of the possibilities of obtaining analytical solutions of DGLAP evolution equations are always interesting. Some approximated analytical solutions of DGLAP evolution equations suitable at low-x, have been reported in recent years [11, 14] with considerable phenomenological success. Among these methods using Taylor expansion [12], method applying Regge behaviour of structure functions [13], method of characteristics [14] etc. are important. The structure functions thus calculated are expected to rise approximately with a power of x towards low-x which is supported by Regge theory [15, 16]. The low-x region of DIS offers a unique possibility to explore the Regge limit [15] of PQCD. The low-x behaviour (at fixed-$Q^2$) of parton distributions can be considered by a triple pole pomeron model [16, 17] at the initial scale $Q_0^2$ and then evolved using DGLAP equations. The Regge behaviour of the sea quark and antiquark distributions is given by qsea(x) ~ $x^{\lambda p}$ with pomeron exchange [16] of intercept λp = –1. But the valence quark distribution for low-x given by qval(x) ~ $x^{-\lambda r}$ corresponding to a reggeon exchange of intercept λr = 1/2. In our present work, we have derived the solutions of singlet and non-singlet DGLAP evolution equations in NLO at low-x limit applying Regge behaviour of structure functions. Here, section 1, section 2, section 3 and section 4 are the introduction, theory, results and discussion, and conclusions respectively.

[1]jamil@tezu.ernet.in, [2]jks@tezu.ernet.in

**2 Theory**

The differential coefficient of singlet structure function $F_2^S(x, Q^2)$ with respect to $\ln Q^2$ i.e. $\partial F_2^S(x, Q^2)/\partial \ln Q$ has a relation with singlet structure function itself as well as gluon distribution function from DGLAP evolution equations [18-20]. The NLO DGLAP evolution equations for singlet and non-singlet structure functions have the standard forms [12]

$$\frac{\partial F_2^S(x,t)}{\partial t} - \frac{\alpha_s(t)}{2\pi}\left[\frac{2}{3}\{3+4\ln(1-x)\}F_2^S(x,t) + \frac{4}{3}\int_x^1 \frac{d\omega}{1-\omega}\left\{(1+\omega^2)F_2^S\left(\frac{x}{\omega},t\right) - 2F_2^S(x,t)\right\}\right.$$

$$+ n_f \int_x^1 \{\omega^2 + (1-\omega)^2\} G\left(\frac{x}{\omega}, t\right) d\omega \Bigg] - \left(\frac{\alpha_s(t)}{2\pi}\right)^2 \Bigg[ (x-1) F_2^S(x,t) \int_0^1 f(\omega) d\omega + \int_x^1 f(\omega) F_2^S\left(\frac{x}{\omega}, t\right) d\omega$$

$$+ \int_x^1 F_{qq}^S(\omega) F_2^S\left(\frac{x}{\omega}, t\right) d\omega + \int_x^1 F_{qg}^S(\omega) G\left(\frac{x}{\omega}, t\right) d\omega \Bigg] = 0 \qquad (1)$$

and

$$\frac{\partial F_2^{NS}(x,t)}{\partial t} - \frac{\alpha_s(t)}{2\eth}\Bigg[\frac{2}{3}\{3 + 4\ln(1-x)\} F_2^{NS}(x,t) + \frac{4}{3}\int_x^1 \frac{d\omega}{1-\omega}\left\{(1+\omega^2) F_2^{NS}\left(\frac{x}{\omega}, t\right) - 2 F_2^{NS}(x,t)\right\}\Bigg]$$

$$- \left(\frac{\alpha_s(t)}{2\eth}\right)^2 \Bigg[(x-1) F_2^{NS}(x,t) \int_0^1 f(\omega) d\omega + \int_x^1 f(\omega) F_2^{NS}\left(\frac{x}{\omega}, t\right) d\omega \Bigg] = 0, \qquad (2)$$

where $F_2^S$ and $F_2^{NS}$ are combinations of quarks and antiquarks, $t = \ln\frac{Q^2}{\Lambda^2}$ and $\alpha_s(t) = \frac{4\pi}{\beta_0 t}\left[1 - \frac{\beta_1 \ln t}{\beta_0^2 t}\right]$, $N_f$ being the number of flavours and $\Lambda$ is the QCD cut-off parameter depends on the renormalization schemes, $â_0$ and $â_1$ are the expansion coefficient of the $â$-function and they are given by $\beta_0 = \frac{33 - 2n_f}{3}$, $\beta_1 = \frac{306 - 38n_f}{3}$. And also

$$f(\omega) = C_F^2 [P_F(\omega) - P_A(\omega)] + \frac{1}{2} C_F C_A [P_G(\omega) + P_A(\omega)] + C_F T_R n_f P_{N_f}(\omega),$$

and

$$F_{qq}^S(\omega) = 2 C_F T_R n_f F_{qq}(\omega), \quad F_{qg}^S(\omega) = C_F T_R n_f F_{qg}^1(\omega) + C_G T_R n_f F_{qg}^2(\omega).$$

The explicit forms of higher order kernels are [12,18-20]

$$P_F(\omega) = -\frac{2(1+\omega^2)}{1-\omega} \ln\omega \ln(1-\omega) - \left(\frac{3}{1-\omega} + 2\omega\right)\ln\omega - \frac{1}{2}(1+\omega)\ln^2\omega - 5(1-\omega),$$

$$P_G(\omega) = \frac{1+\omega^2}{1-\omega}\left(\ln^2\omega + \frac{11}{3}\ln\omega + \frac{67}{9} - \frac{\pi^2}{3}\right) + 2(1+\omega)\ln\omega + \frac{40}{3}(1-\omega),$$

$$P_{N_F}(\omega) = \frac{2}{3}\left[\frac{1+\omega^2}{1-\omega}\left(-\ln\omega - \frac{5}{3}\right) - 2(1-\omega)\right],$$

$$P_A(\omega) = \frac{2(1+\omega^2)}{1+\omega}\int_{\omega/(1+\omega)}^{1/(1+\omega)} \frac{dk}{k}\ln\frac{1-k}{k} + 2(1+\omega)\ln\omega + 4(1-\omega),$$

$$F_{qq}(\omega) = \frac{20}{9\omega} - 2 + 6\omega - \frac{56}{9}\omega^2 + \left(1 + 5\omega + \frac{8}{3}\omega^2\right)\ln\omega - (1+\omega)\ln^2\omega,$$

$$F_{qg}^1(\omega) = 4 - 9\omega - (1-4\omega)\ln\omega - (1-2\omega)\ln^2\omega + 4\ln(1-\omega) + \left[2\ln^2\left(\frac{1-\omega}{\omega}\right) - 4\ln\left(\frac{1-\omega}{\omega}\right) - \frac{2}{3}\pi^2 + 10\right]P_{qg}(\omega)$$

and

$$F_{qg}^2(\omega) = \frac{182}{9} + \frac{14}{9}\omega + \frac{40}{9\omega} + \left(\frac{136}{3}\omega - \frac{38}{3}\right)\ln\omega - 4\ln(1-\omega) - (2+8\omega)\ln^2\omega +$$

$$\left[-\ln^2\omega + \frac{44}{3}\ln\omega - 2\ln^2(1-\omega) + 4\ln(1-\omega) + \frac{\omega^2}{3} - \frac{218}{9}\right]P_{qg}(\omega) + 2P_{qg}(-\omega)\int_{\omega/1+\omega}^{1/1+\omega}\frac{dz}{z}\ln\frac{1-z}{z},$$

where

$P_{qg}(\omega) = \omega^2 + (1-\omega)^2$, and $C_A$, $C_G$, $C_F$, and $T_R$ are constants associated with the color SU(3) group, and $C_A = C_G = N_C = 3$, $C_F = (N_C^2 - 1)/2N_C$ and $T_R = 1/2$. $N_C$ is the number of colours.

Now let us consider the Regge Behaviour of singlet and non-singlet structure functions [16, 17, 21, 22] as

$$F_2^S(x,t) = T_1(t)x^{-\lambda_S} \text{ and } F_2^{NS}(x,t) = T_2(t)x^{-\lambda_{NS}}, \tag{3}$$

where $T_1(t)$ and $T_2(t)$ are functions of $Q^2$ only and $\lambda_S$ and $\lambda_{NS}$ are the Regge intercepts for singlet and non-singlet structure functions respectively. From equation (3) we get

$$F_2^S(x/\omega, t) = T_1(t)\omega^{\lambda_S} x^{-\lambda_S} = \omega^{\lambda_S} F_2^S(x,t) \tag{4}$$

and

$$F_2^{NS}(x/\omega,t) = T_2(t)\omega^{\lambda_{NS}} x^{-\lambda_{NS}} = \omega^{\lambda_{NS}} F_2^{NS}(x,t). \quad (5)$$

Since the DGLAP evolution equations of gluon and singlet structure functions in leading order (LO) and next-to-leading order (NLO) are in the same forms of derivative with respect to t, so we consider the ansatz [13, 14, 23]

$$G(x, t) = K(x) F_2^s(x, t) \quad (6)$$

for simplicity, where K(x) is a parameter to be determined from phenomenological analysis and we assume $K(x) = k$, $ax^b$ or $ce^{dx}$, where k, a, b, c and d are constants. Though we have assumed some simple standard functional forms of K(x), yet we can not rule out the other possibilities. So, we have to consider k, a, b, c and d as some parameters. Actual functional form of K(x) can be determined by simultaneous solution of coupled equations of gluon and singlet structure functions. Therefore

$$G\left(\frac{x}{\omega}, t\right) = K\left(\frac{x}{\omega}\right) F_2^S\left(\frac{x}{\omega}, t\right) = K\left(\frac{x}{\omega}\right) \omega^{\lambda_S} F_2^S(x, t). \quad (7)$$

Putting equations (3), (4) and (7) in equation (1) we arrive at

$$\frac{\partial F_2^S(x,t)}{\partial t} - F_2^S(x,t) \cdot P(x,t) = 0, \quad (8)$$

where

$$P(x, t) = \frac{\alpha_s(t)}{2\pi} \cdot f_1(x) + \left(\frac{\alpha_s(t)}{2\pi}\right)^2 \cdot f_2(x),$$

$$f_1(x) = \frac{2}{3}\{3 + 4\ln(1-x)\} + \frac{4}{3}\int_x^1 \frac{d\omega}{1-\omega}\{(1+\omega^2)\omega^\lambda - 2\} + N_f \int_x^1 \{\omega^2 + (1-\omega)^2\} K\left(\frac{x}{\omega}\right)\omega^\lambda d\omega$$

and

$$f_2(x) = (x-1)\int_0^1 f(\omega)d\omega + \int_x^1 f(\omega)\omega^\lambda d\omega + \int_x^1 F_{qq}^S(\omega)\omega^\lambda d\omega + \int_x^1 F_{qg}^S(\omega) K\left(\frac{x}{\omega}\right)\omega^\lambda d\omega.$$

For possible solutions in NLO, we have taken the expression for $\left(\frac{\alpha_s(t)}{2\pi}\right)$ upto LO correction and we have to put an extra assumption $\left(\frac{\alpha_s(t)}{2\pi}\right)^2 = T_0\left(\frac{\alpha_s(t)}{2\pi}\right)$ [12, 20], where $T_0$ is a numerical parameter. But $T_0$ is not arbitrary. We choose $T_0$ such that difference between $T^2(t)$ and $T_0 T(t)$ is minimum in the region of our discussion. (see fig.3(b)). Now equation (8) reduces to

$$\frac{\partial F_2^S(x,t)}{\partial t} - \frac{F_2^S(x,t)}{t} \cdot P(x) = 0, \tag{9}$$

where

$$P(x) = \frac{2}{\beta_0} \cdot f_1(x) + T_0 \cdot \frac{2}{\beta_0} \cdot f_2(x).$$

Integrating equation (9) we get

$$F_2^S(x,t) = C t^{P(x)}, \tag{10}$$

where C is a constant of integration. This gives the singlet structure function derived by solving NLO DGLAP evolution equation applying Regge behaviour of singlet structure function.

Pursuing the same procedure we get from equation (2)

$$F_2^{NS}(x,t) = C t^{Q(x)}, \tag{11}$$

where

$$Q(x) = \frac{2}{\beta_0} \cdot f_3(x) + T_0 \cdot \frac{2}{\beta_0} \cdot f_4(x)$$

$$f_3(x) = \frac{2}{3}\{3 + 4\ln(1-x)\} + \frac{4}{3}\int_x^1 \frac{d\omega}{1-\omega}\left\{(1+\omega^2)\omega^\lambda - 2\right\}$$

and

$$f_4(x) = (x-1)\int_0^1 f(\omega)d\omega + \int_x^1 f(\omega)\omega^\lambda d\omega.$$

This gives the non-singlet structure function derived by solving NLO DGLAP evolution equation applying Regge behaviour of non-singlet structure function.

For phenomenological analysis, we compare our results with various experimental structure functions. Deuteron and proton structure functions [2, 24] can be written in terms of singlet and non-singlet quark distribution functions as

$$F_2^d(x,t) = \frac{5}{9} F_2^S(x,t) \tag{12}$$

and

$$F_2^p(x,t) = \frac{3}{18} F_2^{NS}(x,t) + \frac{5}{18} F_2^S(x,t). \tag{13}$$

Applying initial conditions at $x = x_0$, $F_2^S(x,t) = F_2^S(x_0,t)$, $F_2^{NS}(x,t) = F_2^{NS}(x_0,t)$ and at $t = t_0$, $F_2^S(x,t) = F_2^S(x,t_0)$, $F_2^{NS}(x,t) = F_2^{NS}(x,t_0)$, we found the t and x-evolution equations for the deuteron and proton structure functions respectively as

$$F_2^d(x,t) = F_2^d(x,t_0)\left(\frac{t}{t_0}\right)^{P(x)}, \tag{14}$$

$$F_2^d(x,t) = F_2^d(x_0,t) t^{\{P(x) - P(x_0)\}}, \tag{15}$$

$$F_2^P(x,t) = F_2^P(x,t_0) \frac{3t^{Q(x)} + 5t^{P(x)}}{3t_0^{Q(x)} + 5t_0^{P(x)}} \tag{16}$$

and

$$F_2^P(x,t) = F_2^P(x_0,t) \frac{3t^{Q(x)} + 5t^{P(x)}}{3t^{Q(x_0)} + 5t^{P(x_0)}}. \tag{17}$$

## 3 Results and discussion

We obtained a new description of t and x-evolutions of deuteron and proton structure functions in NLO considering Regge behaviour of singlet and non-singlet structure functions at low-x. We compare our result of deuteron (proton) structure function with the data set measured by the NMC [25] in muon-deuteron DIS from the merged data sets at incident momenta 90, 120, 200 and 280 GeV$^2$ and also with the data set measured by the Fermilab E665 [26] Collaboration in muon-deuteron DIS at an average beam energy of 470 GeV$^2$. Data cover the x range 0.0008 to 0.6 and $Q^2$ range from 0.2 to 75 GeV$^2$. Here we take the QCD cut-off parameter $\Lambda_{\overline{MS}}(N_f = 4) = 323$ MeV for $\alpha_s(M_z^2) = 0.119 \pm 0.002$ [27]. Deuteron and proton structure functions measured in the range of $0.75 \leq Q^2 \leq 9.795$ GeV$^2$, $0.0045 \leq x \leq 0.0173$ and in the range of $18.323 \leq Q^2 \leq 27$ GeV$^2$, $0.04898 \leq x \leq 0.11$ have been used for phenomenological analysis of t and x-evolutions of these structure functions in NLO.

The comparisons of our results with experimental data sets are made for $\lambda_S = \lambda_{NS} =$ constant. As the value of $\lambda_S$ and $\lambda_{NS}$ should be close to 0.5 in a quite broad range of low-x [13, 16, 21, 28], we have taken $\lambda_S = \lambda_{NS} = 0.5$. The best fit results were found in the range of our discussion. We compare our results for $K(x) = k$, $ax^b$ and $ce^{dx}$, where k, a, b, c and d are constants. But agreement of the results of t and x-evolutions of proton structure function with experimental data is found to be very poor for $K(x) = k$ and $ce^{dx}$. So we present only the results with K(x) =ax$^b$ for proton structure functions. Our result of t evolution of deuteron structure function is also very poor for for K(x) = k and ce$^{dx}$. So, for this evolution also we present only the result with K(x) =ax$^b$. And our result of t evolution of deuteron structure function is also very poor for K(x) = k. Therefore we do not present our result with K(x) =k for t evolution of deuteron structure function.

In fig.1(a-b), we present our result of t-evolution of deuteron structure function (solid lines) for the representative values of x in NLO. Data points at lowest-$Q^2$ values in the figures are taken as input to test the evolution equation. Agreement with the data for $\lambda_S$ = 0.5, 21≤a≤57 and b=2 is good. We observe that when x increases the value of K(x) decreases. In Figure 1(c-d) we present our result for x-evolution of deuteron structure function (solid lines) for the representative values of $Q^2$ in NLO. The best fit results were found for $\lambda_d$ = 0.5, 1≤a≤1.8, b=1, 0.8≤c≤1.4, and d=1 in the x-$Q^2$ range of our discussion with the data. In Figure 2(a-b), we present our results for t-evolution of proton structure function (solid lines) for the representative values of x in NLO. Agreement with the data for $\lambda_S$=$\lambda_{NS}$= 0.5, 30≤a≤63 and b=2 is good. In Figure 2(c-d), we present our results for x-evolution of proton structure function (solid lines) for the representative values of $Q^2$ in NLO. Agreement with the data for $\lambda_S$=$\lambda_{NS}$= 0.5, 5≤a≤15, b=2, in the x-$Q^2$ range of our discussion. Fig. 3(a) shows our best fit graphs for both LO and NLO results for x-evolution of deuteron structure function with NMC data. In case of LO the best fitted results are obtained at k=7, a=7, b=0.001, c=10, d=0.1 for $Q^2$=20 GeV$^2$ and at k=6.5, a=6.5, b=0.001 ,c=8.5, d=0.1 for $Q^2$=27 GeV$^2$. In case of NLO, best fitted results are obtained at a=1, b=1, c=0.8, d=1 for $Q^2$=20 GeV$^2$ and at a=1.05, b=1, c=0.85, d=1 for $Q^2$=27 GeV$^2$. We observe that x-evolutions show more power behaviour in NLO result than those of LO. Therefore it is obvious that agreement with the NLO results is better than with the LO results. In Fig. 3(b) we plot T(t)$^2$ and T$_0$T(t), where T(t) = á$_s$(t)/2ð against $Q^2$ in the $Q^2$ range 0 ≤ $Q^2$≤ 30 GeV$^2$ as required by our data used. Here we observe that for T$_0$ = 0.108, errors become minimum in the $Q^2$ range of our discussion 0.75 ≤ $Q^2$≤ 27 GeV$^2$. The difference between the values of T(t)$^2$ and T$_0$T(t) in this range comes out nearly around 0.28% which is negligible. In fig.4 (a-e), we present the sensitivity of our results for T$_0$, $\lambda_S$, a, b, c and d in NLO with the data set of NMC for the x-evolution of deuteron structure function. If the values of a, c, or d respectively are increased, the curves shift upward and if the values of T$_0$, a, c, or d respectively are decreased, the curves move in the opposite direction. On the other hand if values of $\lambda_S$ or b increased or decreased the curve goes downward or upward directions respectively. We found the ranges of the parameters as $0.128 \leq T_0 \leq 0.088$, $0.4 \leq \lambda_S \leq 0.6$, $1.1 \leq a \leq 0.9$, $1.15 \leq b \leq 0.85$, $0.87 \leq c \leq 0.73$ and $1.1 \leq d \leq 0.9$.

**4 Conclusions**

In our present work, we have considered the Regge behaviour of singlet and non-singlet structure functions to solve DGLAP evolution equations. Here we find the t and x-evolutions of deuteron and proton structure functions in NLO. We see that our results are in good agreement with New Muon and E665 collaborations data sets especially at low-x and high-$Q^2$ region. We can conclude that Regge behaviour of quark is compatible with PQCD at that region. Though we have simplified our solution through a numerical variable $T_0$, yet we have not taken the value arbitrarily. The value has been chosen in such a manner that difference between $T^2(t)$ and $T_0 T(t)$ is negligible in the region of our discussion. Considering Regge behaviour of distribution functions DGLAP equations become quite simple to solve and so this method is a viable simple alternative to other methods. But here also the problem of ad hoc assumption of the function K(x), the relation between singlet structure function and gluon distribution function, could not be overcome. It can be done by the simultaneous solution of coupled DGLAP evolution equations for singlet structure function and gluon distribution function and it has been already done in LO [29]. Again in our solution, the number of parameters used is also less compared to other standard methods. Moreover, the ranges of values of the parameters used are also narrow.

**Figure captions**

**Fig. 1.** t and x-evolutions of deuteron structure function in NLO for the representative values of x and $Q^2$. Data points at lowest-$Q^2$ values are taken as input to test the evolution equation (14) and data points for x values just below 0.1 are taken as input to test the evolution equation (15). Here Fig. 1(a)-1(b) are the best fit graphs of our result of t-evolution for $\lambda_d =$

0.5 and K(x) = $ax^b$ with NMC and E665 data. And Fig. 1(c)-1(d) are the best fit graphs of our result of x-evolution for $\lambda_d$ = 0.5 and K(x) = $ax^b$ and $ce^{dx}$ with NMC and E665 data.

**Fig. 2.** t and x-evolutions of proton structure function in NLO for the representative values of x and $Q^2$. Data points at lowest-$Q^2$ values are taken as input to test the evolution equation (16) and data points for x values just below 0.1 are taken as input to test the evolution equation (17). Here Fig. 2(a)-2(b) are the best fit graphs of our result of t-evolution for $\lambda_S$ = $\lambda_{NS}$ =0.5 with NMC and E665 data. And Fig. 2(c)-2(d) are the best fit graphs of our result of x-evolution for $\lambda_d$ = 0.5 with NMC and E665 data.

**Fig. 3.** Fig. 3(a) Shows both our best fit graphs of LO and NLO results for x-evolution of deuteron structure function with NMC data. Fig. 3(b) shows the variation of $T(t)^2$ and $T_0T(t)$ with $Q^2$.

**Fig. 4.** Fig. 4(a)-4(f) show the sensitivity of the parameters $T_0$, $\lambda$, a, b, c and d respectively at $Q^2$ = 20 $GeV^2$ with the best fit graph of our results with NMC data.

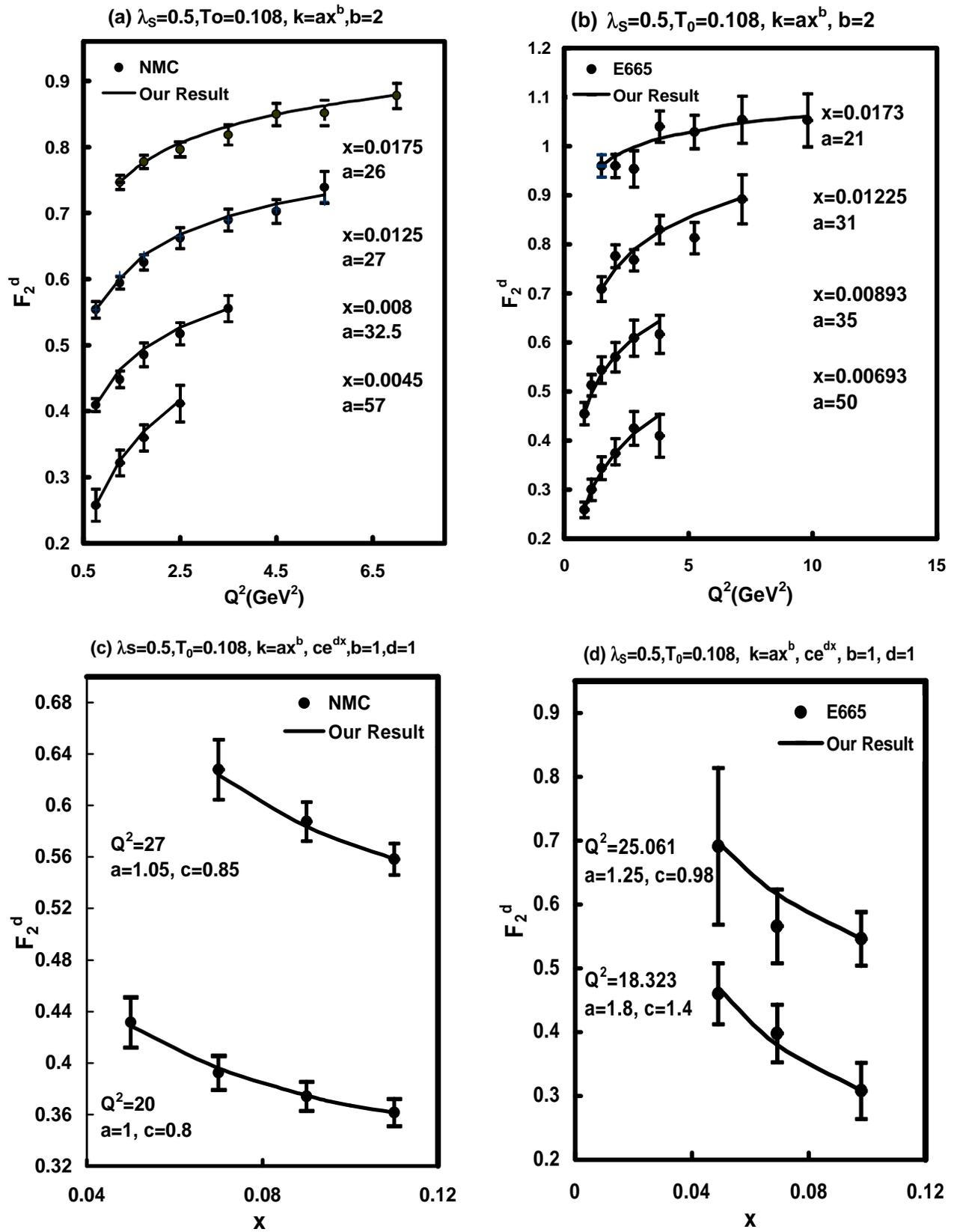

**Fig. 1**

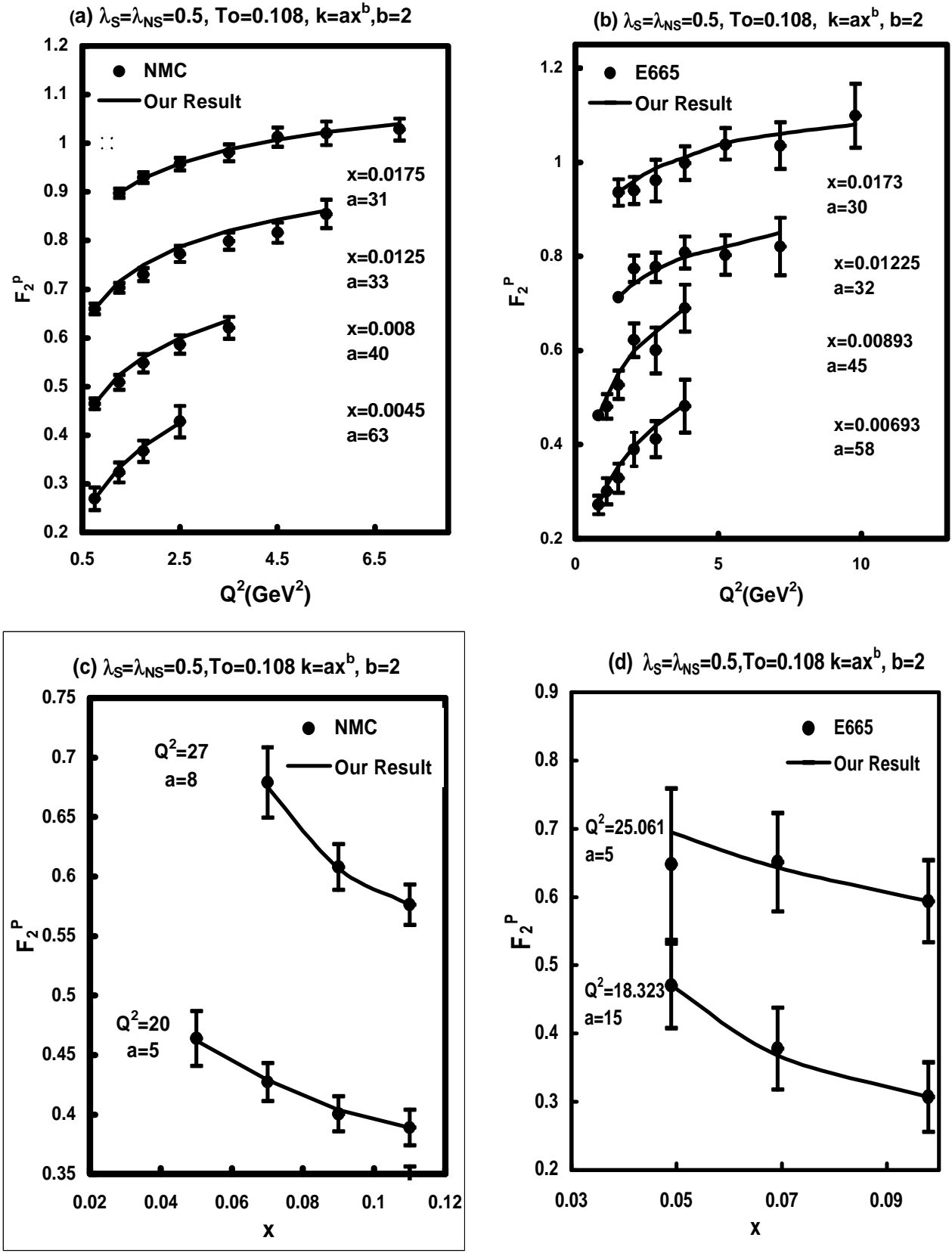

**Fig. 2**

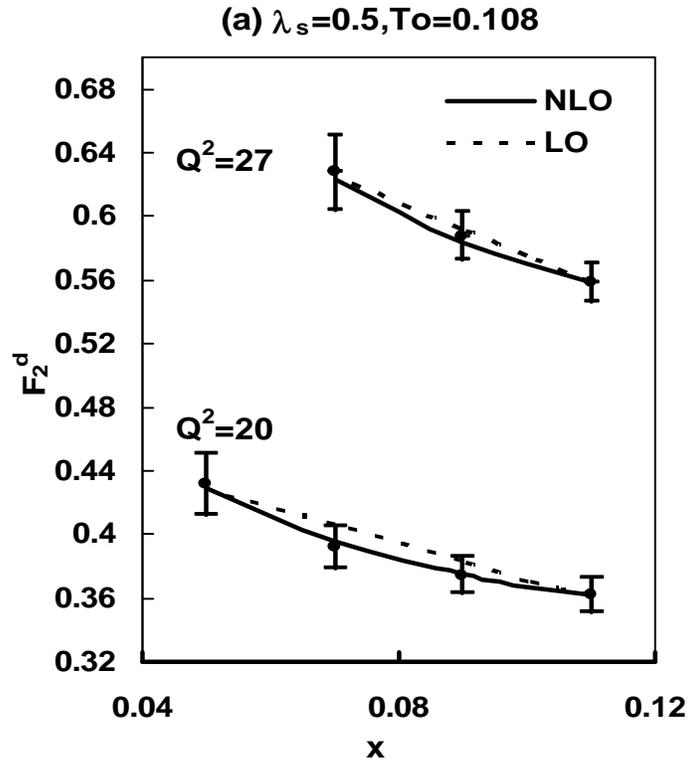

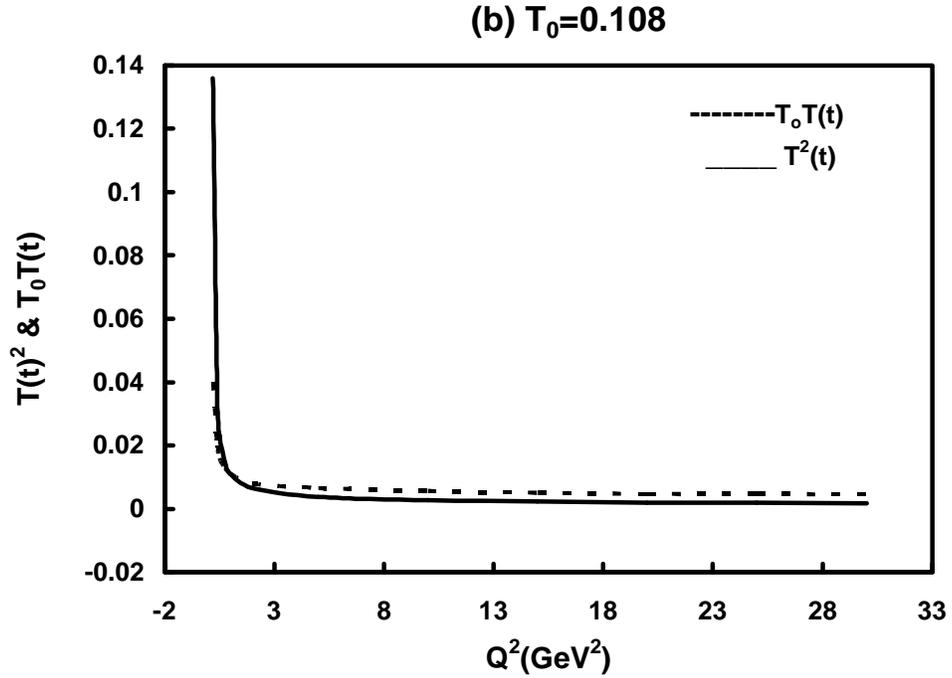

**Fig.3**

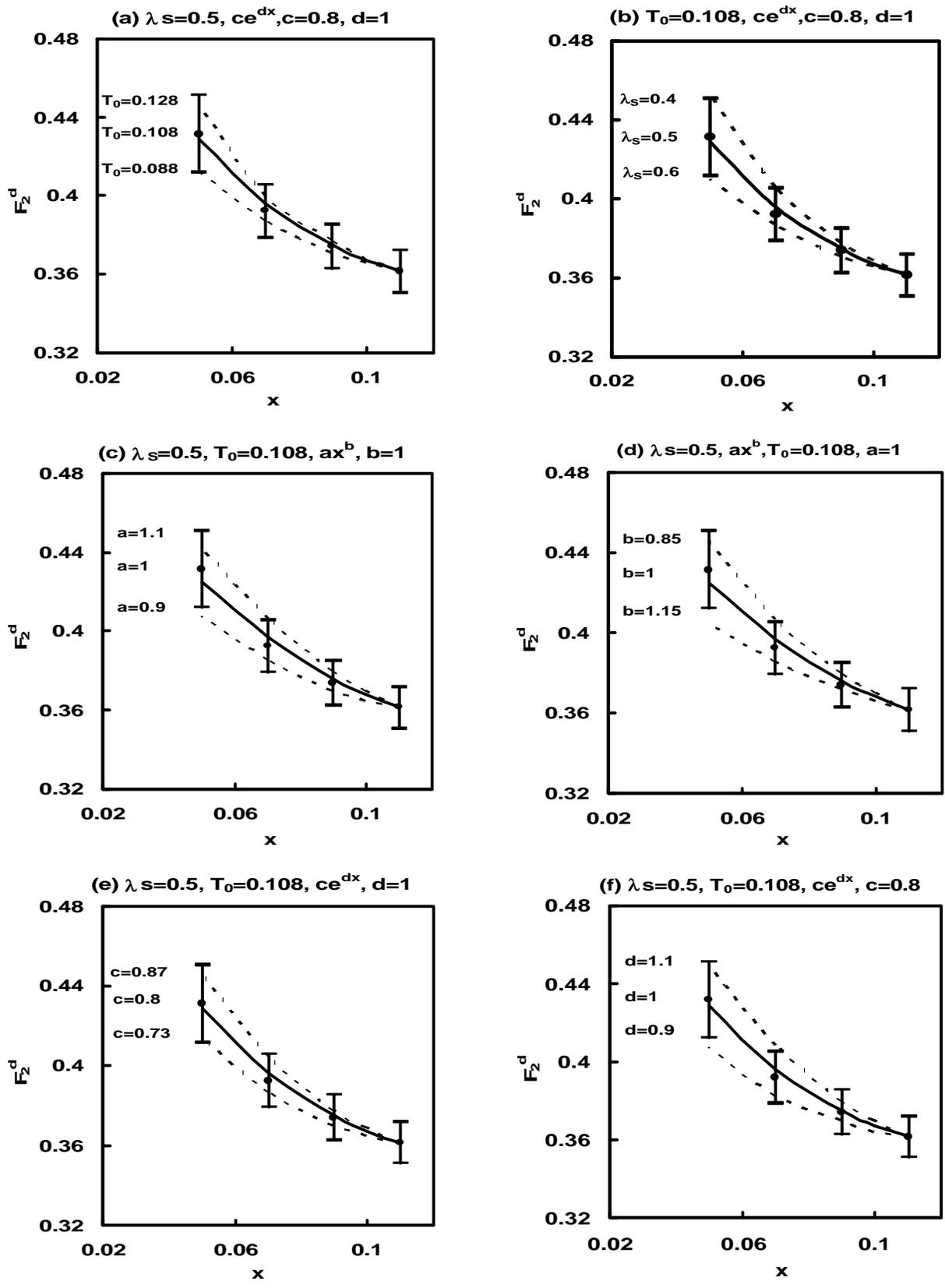

**Fig.4**